\documentclass[12pt,twoside]{article}
\usepackage[mathscr]{eucal}
\usepackage{amsmath,amsfonts,amssymb,amsthm}
\usepackage[matrix,arrow]{xy}
\usepackage{times}
\usepackage{graphicx}
\usepackage{epstopdf}
\usepackage{multibox}
\usepackage{dcolumn}
\usepackage{hyperref}

\def\d_Vphi{\text{d}_V\hspace{-0.06em}\phi}
\def\d_Vphibar{\text{d}_V\hspace{-0.06em}\bar\phi}
\def\d_Vxi{\text{d}_V\hspace{-0.06em}\xi}

\def\be{\begin{eqnarray}}
\def\ee{\end{eqnarray}}
\def\beann{\begin{eqnarray*}}
\def\eeann{\end{eqnarray*}}
\def\beq{\begin{equation}}
\def\eeq{\end{equation}}
\def\ba{\begin{array}}
\def\ea{\end{array}}
\def\ben{\begin{enumerate}}
\def\een{\end{enumerate}}
\def\bea{\begin{eqnarray}}
\def\eea{\end{eqnarray}}

\def\5{\bar }
\def\6{\partial }
\def\7{\hat }
\def\4{\tilde }
\advance\textheight\topskip
\renewcommand{\tilde}{\widetilde}
\renewcommand{\hat}{\widehat}

\newcommand{\T}{\mathrm{T}}

\newcommand{\dd}{\partial}
\renewcommand{\d}{\partial}

\newcommand{\binner}[2]{%
  {\langle}\kern-4.15pt{\langle}#1{,}\,#2{\rangle}\kern-4.15pt{\rangle}}

\newcommand{\half}{\mathchoice{%
    \ffrac{1}{2}}{\frac{1}{2}}{\frac{1}{2}}{\frac{1}{2}}}

\newcommand{\ffrac}[2]{\raisebox{.5pt}%
  {\footnotesize$\displaystyle\frac{#1}{#2}$}\kern1pt}

\newcommand{\dover}[2]{\ffrac{\dd #1}{\dd #2}}

\def\cI{\mathcal{I}}

\def\cL{\mathcal{L}}

\begin{document}

\def\mytitle{Symmetries of asymptotically flat 4 dimensional spacetimes at null
  infinity revisited}

\pagestyle{myheadings} \markboth{\textsc{\small Barnich, Troessaert}}{%
  \textsc{\small BMS algebra revisited}} \addtolength{\headsep}{4pt}

\begin{flushright}\small
ULB-TH/09-24\end{flushright}

\begin{centering}

  \vspace{1cm}

  \textbf{\Large{\mytitle}}



  \vspace{1.5cm}

  {\large Glenn Barnich$^{a}$ and C\'edric Troessaert$^{b}$}

\vspace{.5cm}

\begin{minipage}{.9\textwidth}\small \it \begin{center}
   Physique Th\'eorique et Math\'ematique\\ Universit\'e Libre de
   Bruxelles\\ and \\ International Solvay Institutes \\ Campus
   Plaine C.P. 231, B-1050 Bruxelles, Belgium \end{center}
\end{minipage}

\vspace{1cm}


\end{centering}

\vspace{1cm}

\begin{center}
  \begin{minipage}{.9\textwidth}
    \textsc{Abstract}. It is shown that the symmetry algebra of
    asymptotically flat spacetimes at null infinity in 4 dimensions
    should be taken as the semi-direct sum of supertranslations with
    infinitesimal local conformal transformations and not, as usually
    done, with the Lorentz algebra. As a consequence, two dimensional
    conformal field theory techniques will play as fundamental a role
    in this context of direct physical interest as they do in three
    dimensional anti-de Sitter gravity.
  \end{minipage}
\end{center}

\vfill

\noindent
\mbox{}
\raisebox{-3\baselineskip}{%
  \parbox{\textwidth}{\mbox{}\hrulefill\\[-4pt]}}
{\scriptsize$^a$Research Director of the Fund for Scientific
  Research-FNRS. E-mail: gbarnich@ulb.ac.be \\$^b$ Research Fellow of
  the Fund for Scientific Research-FNRS. E-mail: ctroessa@ulb.ac.be}

\thispagestyle{empty}
\newpage



In the study of gravitational waves in the early sixties
\cite{Bondi:1962px,Sachs:1962wk}, a lot of efforts have been devoted
to specifying both local coordinate and global boundary conditions at
null infinity that characterize asymptotically flat 4 dimensional
spacetimes. The group of non singular transformations leaving these
conditions invariant is the well-known Bondi-Metzner-Sachs group. It
consists of the semi-direct product of the group of globally defined
conformal transformations of the unit $2$-sphere, which is isomorphic
to the orthochronous homogeneous Lorentz group, times the abelian
normal subgroup of so-called supertranslations.

What seems to have been largely overlooked so far is the fact that,
when one focuses on infinitesimal transformations and does not require
the associated finite transformations to be globally well-defined, the
symmetry algebra is the semi-direct sum of the infinitesimal local
conformal transformations of the $2$-sphere with the abelian ideal of
supertranslations, and now both factors are infinite-dimensional. This
is already obvious from the details of the derivation of the
asymptotic symmetry algebra by Sachs in 1962 \cite{Sachs2}.

Let $x^0=u,x^1=r,x^2=\theta,x^3=\phi$ and $A,B,\dots=2,3$. Following
\cite{Sachs2} up to notation, the metric $g_{\mu\nu}$ of an
asymptotically flat spacetime can be written in the form
\begin{equation}
  \label{eq:2}
  ds^2=e^{2\beta}\frac{V}{r} du^2-2e^{2\beta}dudr+
g_{AB}(dx^A-U^Adu)(dx^B-U^Bdu)
\end{equation}
where $\beta,V,U^A,g_{AB} (\text{det}\, g_{AB})^{-1/2}$ are $6$
functions of the coordinates, with $\text{det}\, g_{AB}=r^4 b$ for a
function $b(u,\theta,\phi)$.  Sachs fixes $b=\sin^2\theta$, but the
geometrical analysis by Penrose \cite{PhysRevLett.10.66} suggests to
keep it arbitrary throughout the analysis.  In order to streamline the
derivation below, it turns out convenient to use the parametrization
$|b|=\frac{1}{4} e^{4\tilde\varphi}$, which implies in particular that
$g^{AB}\d_\alpha
g_{AB}=\d_\alpha\ln{(\frac{r^4}{4}e^{4\tilde\varphi}})$.

The fall-off conditions for $g_{AB}$ are
\begin{equation}
  \label{eq:1}
  g_{AB}dx^Adx^B=r^2\bar \gamma_{AB}dx^Adx^B+O(r),
\end{equation}
where the $2$-dimensional metric $\bar\gamma_{AB}$ is conformal to the
metric of the unit $2$-sphere,
$\bar\gamma_{AB}=e^{2\varphi}{}_0\gamma_{AB}$ and ${}_0
\gamma_{AB}dx^Adx^B=d\theta^2+\sin^2\theta d\phi^2$.  In terms of the
standard complex coordinates $\zeta= e^{i\phi}\cot{\frac{\theta}{2}}$,
the metric on the sphere is conformally flat, $ d\theta^2+\sin^2\theta
d\phi^2=P^{-2}d\zeta d\bar\zeta$,
$P(\zeta,\bar\zeta)=\half(1+\zeta\bar\zeta)$. We thus have
$\bar\gamma_{AB}dx^Adx^B=e^{2\tilde\varphi}d\zeta d\bar\zeta$ with
$\tilde \varphi=\varphi-\ln P$. In the following we denote by $\bar
D_A$ the covariant derivative with respect to $\bar \gamma_{AB}$ and
by $\bar\Delta$ the associated Laplacian.

In the general case, the remaining fall-off conditions
are
\begin{equation}
  \label{eq:3}
\beta=O(r^{-2}), \quad 
U^A=O(r^{-2}),\quad 
V/r=-2r\d_u \tilde \varphi+ \bar \Delta \tilde \varphi +O(r^{-1}).
\end{equation}

The transformations that leave the form of the metric \eqref{eq:2}
invariant up to a conformal rescaling of $g_{AB}$, i.e., up to a shift
of $\tilde \varphi$ by $\tilde\omega(u,x^A)$, are generated by
spacetime vectors satisfying
\begin{equation}
\begin{gathered}
  \label{eq:4aext}
  \cL_\xi g_{rr}=0,\quad \cL_\xi g_{rA}=0,\quad g^{AB}\cL_\xi g_{AB}
  =4\tilde\omega,\\
\cL_\xi
g_{ur}=O(r^{-2}),\quad \cL_\xi g_{uA}=O(1),\quad  \cL_\xi
g_{AB}=O(r),\\
\cL_\xi g_{uu}=-2r\d_u\tilde\omega -2\tilde\omega \bar\Delta
\tilde\varphi +\bar\Delta \tilde \omega + O(r^{-1}). 
\end{gathered}
\end{equation}
The general solution to these equation is
\begin{gather}
  \label{eq:26}
\left\{\begin{array}{l}
  \xi^u=f,\\
\xi^A=Y^A+I^A, \quad  I^A=- f_{,B} \int_r^\infty dr^\prime(
e^{2\beta} g^{AB}),\\
\xi^r=-\half r (\bar D_A \xi^A-f_{,B}U^B+2f\d_u\tilde \varphi-2\tilde
\omega),\quad 
\end{array}\right.
\end{gather}
with $\d_r f=0=\d_r Y$. 
In addition, 
\begin{equation}
\d_u f =f\d_u\tilde \varphi+\half \psi-\tilde\omega\iff f=e^{\tilde \varphi}\big[T+
 \half\int_0^udu^\prime
e^{-\tilde \varphi}(\psi-2\tilde\omega)\big],\label{eq:44ter}
\end{equation}
where we use the notation $\psi= \bar D_A Y^A$ and where $\d_u
T=0=\d_u Y^A$. Finally $Y^A$ is required to be a conformal Killing
vector of $\bar\gamma_{AB}$.

The Lie algebra $\mathfrak{bms}_4$ can be defined as the
semi-direct sum of the Lie algebra of conformal Killing vectors
$Y^A\dover{}{x^A}$ of the Riemann sphere with the abelian ideal
consisting of functions $T(x^A)$ on the Riemann sphere. The bracket is
defined through $(\hat Y,\hat T)=[(Y_1,T_1),(Y_2,T_2)]$
\begin{equation}
\label{eq:32}
\begin{split}
\hat Y^A&= Y^B_1\d_B
Y^A_2-Y^B_1\d_B Y^A_2,\\
\hat T&=Y^A_1\d_A
  T_2-Y^A_2\d_A T_1 +\half (T_1\d_AY^A_2-T_2\d_AY^A_1).
\end{split}
\end{equation}

Consider then the modified Lie bracket
\begin{equation}
  \label{eq:43}
  [\xi_1,\xi_2]_M=[\xi_1,\xi_2]-\delta^g_{
    \xi_1}\xi_2+\delta^g_{ \xi_2}\xi_1,
\end{equation}
where $\delta^g_{\xi_1}\xi_2$ denotes the variation in $\xi_2$
under the variation of the metric induced by $\xi_1$, $\delta^g_{
  \xi_1}g_{\mu\nu}=\cL_{\xi_1}g_{\mu\nu}$.

Let $\cI$ be the real line times the Riemann sphere  with coordinates
$u,x^A=(\zeta,\bar\zeta)$. On $\cI$, consider the scalar field $\tilde
\varphi,\tilde\omega$ and the vectors fields $\bar\xi(\tilde
\varphi,\tilde\omega,T,Y)=f\dover{}{u} +Y^A\dover{}{x^A}$, with $f$ given in
\eqref{eq:44ter} and $Y^A$ an $u$-independent conformal Killing vector
of the Riemann sphere. 

{\em When equipped with the modified bracket, both the vector fields
  $\bar\xi$ and the spacetime vectors \eqref{eq:26} provide a faithful
  representation of the direct sum of $\mathfrak{bms}_4$ with the
  abelian algebra of conformal rescalings, i.e., the space of elements
  of the form $(Y,T,\omega)$ where
  $[(Y_1,T_1,\tilde\omega_1),(Y_2,T_2,\tilde\omega_2)]=(\hat Y,\hat
  T,\widehat{\tilde\omega})$, with $\hat Y,\hat T$ as before and
  $\widehat{\tilde\omega}=0$.}

Depending on the space of functions under consideration, there are
then basically two options which define what is actually meant by
$\mathfrak{bms}_4$.

The first choice consists in restricting oneself to globally
well-defined transformations on the unit or, equivalently, the Riemann
sphere. This singles out the global conformal transformations, also
called projective transformations, and the associated group is
isomorphic to $SL(2,\mathbb C)/\mathbb Z_2$, which is itself
isomorphic to the proper, orthochronous Lorentz group. Associated with
this choice, the functions $T(\theta,\phi)$, which are the generators
of the so-called supertranslations, have been expanded into spherical
harmonics. This choice has been adopted in the original work by Bondi,
van der Burg, Metzner and Sachs and followed ever since, most notably
in the work of Penrose and Newman-Penrose
\cite{PhysRevLett.10.66,newman:863}. A lot of attention has been
devoted to the conformal rescalings and the ``edth'' operator together
with spin-weighted spherical harmonics have been introduced. After
attempts to cut this version of the BMS group down to the standard
Poincar\'e group, it has been taken seriously as an invariance group
of asymptotically flat spacetimes. Its consequences have been
investigated, but we believe that it is fair to say that this version
of the BMS group has had only a limited amount of success.

The second choice that we would like to advocate here is motivated by
exactly the same considerations that are at the origin of the
breakthrough in two dimensional conformal quantum field theories
\cite{Belavin:1984vu}. It consists in focusing on local properties and
allowing the set of all, not necessarily invertible holomorphic
mappings. In this case, Laurent series on the Riemann sphere are
allowed. The general solution to the conformal Killing equations is
$Y^\zeta=Y^\zeta(\zeta)$, $Y^{\bar\zeta}=Y^{\bar\zeta}(\bar\zeta)$ and
the standard basis vectors are choosen as
\begin{equation}
l_n=-\zeta^{n+1}\frac{\d}{\d\zeta},\quad \bar l_n=-\bar
\zeta^{n+1}\frac{\d}{\d\bar \zeta},\quad n\in \mathbb Z\label{eq:55}
\end{equation}
At the same time, let us choose to expand the generators of the supertranslations 
with respect to
\begin{equation}
  T_{m,n}=\zeta^m\bar\zeta^n, 
\quad m,n\in\mathbb Z. \label{eq:15}
\end{equation}
In terms of the basis vectors $l_l\equiv (l_l,0)$ and
$T_{mn}\equiv(0,T_{mn})$, the commutation relations for the complexified
$\mathfrak{bms}_4$ algebra read
\begin{equation}
\boxed{
\begin{gathered}
  \label{eq:37}
[l_m,l_n]=(m-n)l_{m+n},\quad [\bar l_m,\bar l_n]=(m-n)\bar
  l_{m+n},\quad [l_m,\bar l_n]=0, \\
[l_l,T_{m,n}]=(\frac{l+1}{2}-m)T_{m+l,n},
\quad [\bar l_l,T_{m,n}]= (\frac{l+1}{2}-n)T_{m,n+l}. 
\end{gathered}}
\end{equation}
The complexified Poincar\'e algebra is the subalgebra 
spanned by the generators
\begin{equation}
  l_{-1},\,l_0,\,l_1,\quad \bar l_{-1},\, \bar l_0,\, \bar l_1,\quad
T_{0,0},\,T_{1,0},\T_{0,1},\,T_{1,1}.\label{eq:61}
\end{equation}

The considerations above apply for all choices of $\tilde\varphi$
which is freely at our disposal. In the original work of Bondi, van
der Burg, Metzner and Sachs, and in much of the subsequent work, the
choice $\tilde\varphi=-\ln P$ was preferred. From the conformal point
of view, the choice $\tilde\varphi=0$ is interesting as it turns
$\bar\gamma_{AB}$ into the standard flat metric on the Riemann sphere.

The consequences of local conformal invariance need to be taken into
account when studying representations and our result means that two
dimensional conformal field theory techniques should play a major role
both in the classical and quantum theory of gravitational
radiation. For instance, the representation theory based on the
standard BMS group has been discussed in \cite{mccarthy:1992b} and
references therein, while related holographic considerations have
appeared in \cite{Arcioni:2003td,Arcioni:2003xx}. Furthermore,
implications of the supertranslations in the context of asymptotic
quantization \cite{Ashtekar:1981sf,Ashtekar:1987tt} have already been
investigated. It should prove most interesting to extend these
considerations to include the local conformal transformations.

A new perspective also arises for the problem of angular momentum in
general relativity \cite{winicour:1980aa} since the factor algebra of
$\mathfrak{bms}_4$ modulo the abelian ideal of infinitesimal
supertranslations is now the infinitedimensional Virasoro algebra
rather than the Lorentz algebra.

Earlier work where the relevance of conformal field theories for
asymptotically flat spacetimes at null infinity has been discussed by
starting out from the correspondence in the (anti-) de Sitter case
includes
\cite{witten:98xx,Susskind:1998vk,Polchinski:1999ry,deBoer:2003vf,%
  Solodukhin:2004gs,Gary:2009mi}. In particular, a symmetry algebra of
the kind that we have derived has been conjectured in
\cite{Banks:2003vp}.

A motivation for our investigation comes from Strominger's derivation
\cite{Strominger:1998eq} of the Bekenstein-Hawking entropy for black
holes that have a near horizon geometry that is locally $AdS_3$. More
recently, a similar analysis has been applied in the case of an
extreme 4-dimensional Kerr black hole \cite{Guica:2008mu}. Our hope is
to make progress along these lines in the non extreme case.  As a
first step, we have computed the behavior of Bondi's news tensor as
well as the mass and angular momentum aspects under local conformal
transformations in \cite{Barnich:2009xx}, where detailed proofs of all
statements of this letter can also be found. The next step consists in
the construction of the surface charges, generators and central
extensions associated to $\mathfrak{bms}_4$.

\section*{Acknowledgements}
\label{sec:acknowledgements}

The authors thank M.~Ba\~nados, G.~Comp\`ere, G.~Giribet,
A.~Gomberoff, M.~Henneaux, C.~Mart\'{\i}nez, R.~Troncoso and
A.~Virmani for useful discussions. This work is supported in parts by
the Fund for Scientific Research-FNRS (Belgium), by the Belgian
Federal Science Policy Office through the Interuniversity Attraction
Pole P6/11, by IISN-Belgium and by Fondecyt project No.~1085322.


\def\cprime{$'$}
\providecommand{\href}[2]{#2}\begingroup\raggedright\endgroup

\end{document}